\documentclass[showpacs,twocolumn,amsmath,amssymb,pra]{revtex4}
\usepackage{txfonts}
\usepackage{amssymb,amsmath}

\usepackage{graphicx}
\usepackage{dcolumn}
\usepackage{bm}

\begin{document}

\title{\bigskip Proposed solid-state Faraday anomalous-dispersion optical filter}

\author{Wei-Bin Lin,$^{1,2}$ Zong-Quan Zhou,$^{1}$ Chuan-Feng Li,$^{1}\footnote{email: cfli@ustc.edu.cn}$ and  Guang-Can Guo$^{1}$}

\affiliation{ $^1$Key Laboratory of Quantum Information, University of Science and Technology of China, CAS, Hefei, 230026, People's Republic of China
\\$^2$Institute of Laser Technology, Hefei University of Technology, Hefei, 230009, People's Republic of China}
\date{\today }

\begin{abstract}
We propose a Faraday anomalous dispersion optical filter (FADOF) based on a rare-earth ion doped crystal. We present theoretical
analyses for the solid-state FADOF transmission. Our theoretical model predicts a maximum transmission efficiency of $71\%$ and a
double-peaked transmission spectrum with a bandwidth of 6 GHz under current experimental conditions. Our proposal may have important
applications in optical communications.
\end{abstract}
\pacs{42.79.Ci 42.25.Bs}

\maketitle
Ultra-narrow-bandwidth with a wide field of view (FOV) optical filters play key roles in quantum-noise-limited performance
and background-limited laser receivers. One typical ultra-narrow-bandwidth filter is the atomic
resonance filter \cite{N.Bloembergen1958}, which has the drawbacks of slow response and low quantum efficiency \cite{J.A.Gelbwachs1988}.
Another device, referred to as the Faraday anomalous dispersion optical filter (FADOF) \cite{B.Yin1991}, is considered as the best ultra-narrow-bandwidth optical
filter. Since FADOFs have the advantages of high transmission, fast response, a wide FOV, high noise rejection, relative insensitivity
to vibrations, and imaging capability \cite{D.J.Dick1991}, they have been widely applied in various fields, especially in laser-based fields such as free-space laser
communication \cite{Tang Junxiong1995}, dye-laser frequency locking \cite{T.Endo1977}, remote sensing \cite{A.Popescu2005} and lidar \cite{H. Chen 1996,Yang Yong2011}.

Conventional FADOFs consist of an atomic vapor cell placed between two crossed polarizers. A homogeneous, constant magnetic field is applied
parallel to the optical axis of the vapor cell \cite{A. Popescu2010}. The Faraday effect \cite{Paul Siddons2009,Richard P. Abel2009}
plays a key role in FADOFs. Briefly, a plane-polarized light beam consists equally of right- and left-circularly-polarized components. Each
component of circular polarization has a different refractive index because of the Zeeman and hyperfine splitting of the energy levels. This results
in a rotation of the direction of plane polarization. If the incident light is outside the absorption bands and its polarization rotation angle approximates to odd
multiples of $\pi/2$, it will pass through the two crossed polarizers. Therefore, the transmission of FADOFs is strongly dependent on the frequency of the incident
light. This type of FADOF was first introduced by Ohman as early as 1956 \cite{Y.Ohman1956}. The theory behind FADOFs had been discussed
previously by Yeh \cite{P.Yeh1982}; it was improved by Yin and Shay to include hyperfine effects \cite{B.Yin1991}. The first experimental demonstration of a FADOF was given
by Dick \emph{et al} \cite{D.J.Dick1991}. Then, different experiments were completed based on both ground-state transitions \cite{H.Chen1993,Yundong Zhang2001}
and excited-state transitions \cite{R.I.Billmers1995,A. Popescu2010}. Both theoretical and experimental studies have been performed using a variety
of atomic species: the working elements include Na \cite{H.Chen1993,Harrell2009}, K \cite{Dressler1996,R.I.Billmers1995}, Rb \cite{D.J.Dick1991,Alessandro Cere2009},
Cs \cite{B.Yin1991,J.Menders1992}, Ca \cite{J.A.Gelbwachs1993}, and Sr \cite{J.A.Gelbwachs1991}. Most of these FADOFs operate in the line-wing profile \cite{Dressler1996,A.Popescu2005},
others operate in the line-center profile \cite{H.Chen1993,Zhilin Hu1998}, and these two transmission types are interchangeable \cite{Yundong Zhang2001}. However,
all of these studies were performed in vapor systems which have a larger Doppler effect and whose operating wavelengths are limited by several hot atomic species' transitions.
These drawbacks limit their tuning range and extension of applications.

In this Brief Report, we present the scheme of a solid-state FADOF based on a rare-earth ion doped crystal. We develop a complete theory describing the transmission of the FADOF.
We simulate the performance of the FADOF based on current experimental conditions through our theory model.
\begin{figure}[tbph]
\begin{center}
\includegraphics[width= 3.5in]{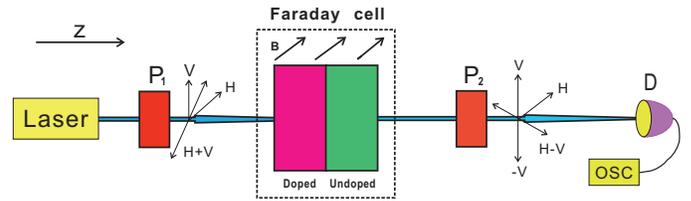}
\end{center}
\caption{(Color online)   Schematic drawing of a solid state FADOF, consisting of a Faraday cell in a homogeneous, constant magnetic field $B$ between two crossed polarizers. $P_{1}$ and $P_{2}$ are the crossed polarizers.
The red and green rectangles denote the doped crystal and undoped crystal, respectively. }
\end{figure}

The solid-state FADOF consists of a Faraday cell in a homogeneous, constant magnetic field $B$, between two crossed polarizers (Fig. 1).
Here we take the Nd$^{3+}$:YVO$_{4}$ FADOF as an example. The Faraday cell of the FADOF is composed of two adjacent crystals. The first one is a Nd$^{3+}$
doped YVO$_{4}$ crystal and the next one is an undoped YVO$_{4}$ crystal, whose c-axis is perpendicular to the $c$ axis of the doped YVO$_{4}$ crystal.
The magnetic field splits the energy levels of the Nd$^{3+}$ ions, which separates the absorption bands into horizontally and
vertically polarized components and introduces a strong anomalous dispersion. A linearly polarized light beam consists of horizontally
and vertically polarized components. Each component of horizontal and vertical polarization has its own refractive index due to the Faraday
anomalous dispersive effect. This results in a polarization rotation when the light beam propagates through the system along the $z$ direction.
When the frequency of incident light is in the vicinity of the Nd$^{3+}$ ions' resonance lines, the Faraday anomalous dispersion causes a strong
frequency dependence of the filter's transmission. Similar to the conventional FADOFs based on atomic vapors, if the incident light is outside the
absorption bands and its polarization rotation angle approximates to odd multiples of $\pi/2$, it will pass through the two crossed polarizers. Therefore, transmission
peaks with a narrow bandwidth can be detected at some specific frequencies.

The incident light is a frequency independent constant $E_{\scriptscriptstyle 0}$. $H$ and $V$ respectively denote the direction parallel and perpendicular to the $c$ axis of the
doped YVO$_{4}$ crystal. $\vec{E_{\scriptscriptstyle H}(z)}$ and $\vec{E_{\scriptscriptstyle V}(z)}$, respectively, denote the light polarized along the $H$
and $V$ directions. The $H+V$ polarized light $\vec{E_{\scriptscriptstyle in}}$ incident on the doped YVO$_{4}$
crystal surface at $z=0$ can be described in terms of horizontally and vertically polarized basis states,
\begin{eqnarray}
\vec{E_{\scriptscriptstyle in}}=\frac{E_{\scriptscriptstyle 0}}{\sqrt{2}}(\vec{e_{\scriptscriptstyle H}}+\vec{e_{\scriptscriptstyle V}})
           =\vec{E_{\scriptscriptstyle H}}(0)+\vec{E_{\scriptscriptstyle V}}(0) \ ,
\end{eqnarray}
where $\vec{e_{\scriptscriptstyle H}}$  and  $\vec{e_{\scriptscriptstyle V}}$  are unit vectors along $H$ and $V$ directions, respectively.
The different refractive indices for $H$- and $V$-polarized light caused by the birefringence effect in the YVO$_{4}$ crystal will lead
to an intrinsic phase delay between the two components at the end of the  doped YVO$_{4}$ crystal. We use an undoped YVO$_{4}$ crystal with
an identical length to compensate for the intrinsic phase delay. Since the $c$ axis of the undoped YVO$_{4}$ crystal is perpendicular to the $c$ axis of the
doped YVO$_{4}$ crystal, the phase delays caused by birefringence can offset each other. With compensation, the net phase delay between the two components is caused by the
Faraday anomalous dispersion effect; the field at the end of the undoped YVO$_{4}$ crystal surface $z=2L$ yields
\begin{eqnarray}
\vec{E_{_{\scriptscriptstyle H/V}}}(2L)=\vec{E_{_{\scriptscriptstyle H/V}}}(0)\ exp\ \big[i\ k_{\scriptscriptstyle H/V}(\omega)L\big] \ ,
\end{eqnarray}
where $L$ is the length of the doped YVO$_{4}$ crystal, $\omega$ is the frequency of incident light, and $k_{\scriptscriptstyle H/V}(\omega)$
is the complex wavenumber for $H$- and $V$-polarized light. After passing the second polarizer, which is placed at the $H-V$ direction, the output field
$\vec{E_{\scriptscriptstyle out}}$ can be expressed as

\begin{eqnarray}
&\vec{E_{\scriptscriptstyle out}}&=\frac{E_{\scriptscriptstyle 0}}{2}\ \Big\{\ exp\ \big[\ i\ k_{\scriptscriptstyle H}(\omega)L\ \big]-exp\ \big[\ i\ k_{\scriptscriptstyle V}(\omega)L\ \big]\ \Big\} \ .
\end{eqnarray}
The frequency-dependent transmission of the Nd$^{3+}$:YVO$_{4}$ FADOF, which is defined by
$T(\omega)=\left|\vec{E_{\scriptscriptstyle out}}\right|^{2}\Big/\ \left|\vec{E_{\scriptscriptstyle in}}\right|^{2}$
, is then given by
\begin{eqnarray}
T(\omega)=&\dfrac{1}{4}&exp\ \Big\{-2Im[k_{\scriptscriptstyle H}(\omega)]L\Big\}+\dfrac{1}{4}\ exp\ \Big\{-2Im[k_{\scriptscriptstyle V}(\omega)]L\Big\} \nonumber \\
&-&\dfrac{1}{2}\ \cos[2\Phi(\omega)]\ exp\ \Big\{Im[k_{\scriptscriptstyle H}(\omega)+k_{\scriptscriptstyle V}(\omega)]L\Big\} \ ,
\end{eqnarray}
Here the Faraday rotation angle of the polarization, $\Phi(\omega)$, is characterized by
\begin{eqnarray}
\Phi(\omega)=\frac{L}{2}\ Re[k_{\scriptscriptstyle H}(\omega)-k_{\scriptscriptstyle V}(\omega)] \ .
\end{eqnarray}
\begin{figure}[tbph]
\begin{center}
\includegraphics[width= 2.8in]{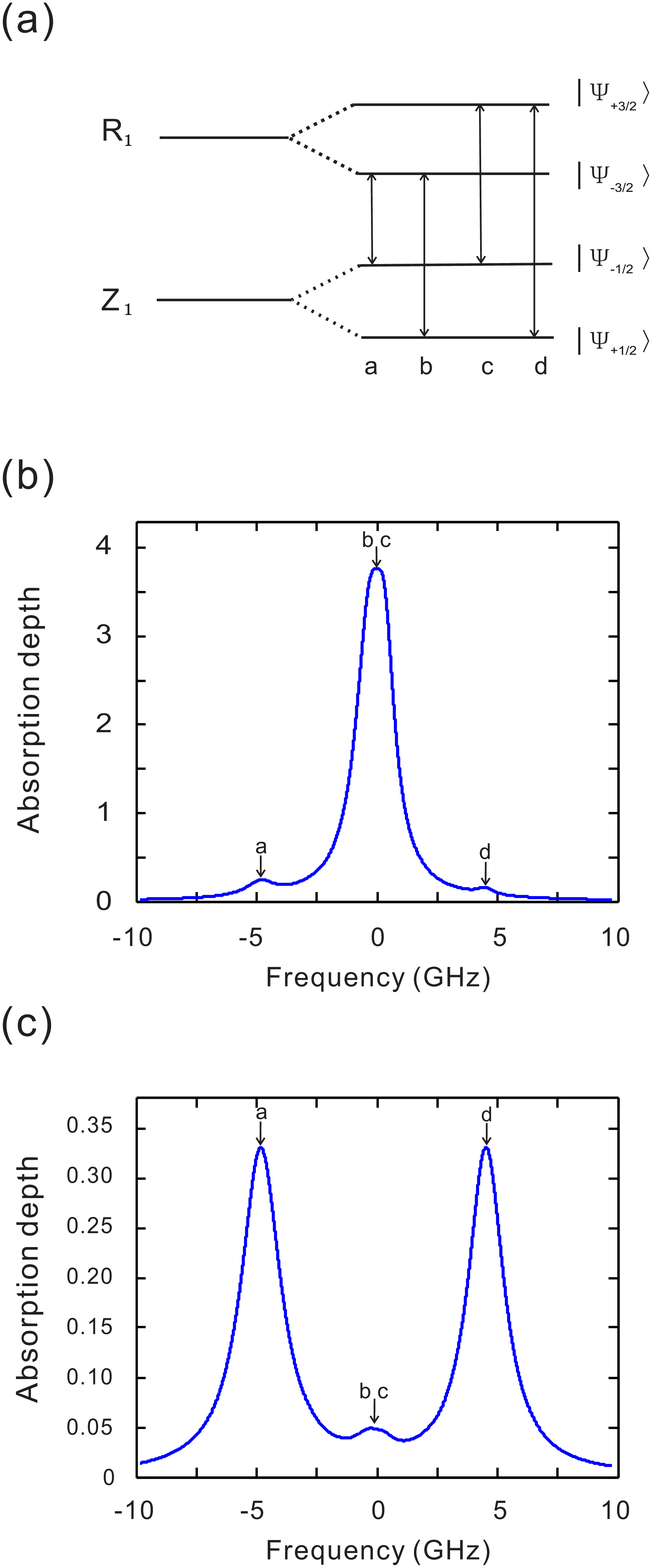}
\end{center}
\caption{(Color online)   (a)  The energy-level diagram of the $^{4}I_{9/2}\leftrightarrow{ }^4F_{3/2}$ transition of Nd$^{3+}$.
(b)  The simulation results of the absorption depth $d_{\scriptscriptstyle H}(\omega)$ as a function of $\omega$ for $H$-polarized light.
(c)  The simulation results of the absorption depth $d_{\scriptscriptstyle V}(\omega)$ as a function of $\omega$ for $V$-polarized light.
In these simulations, the magnetic field $B=0.31\ T$ and the crystal length $L=0.9\ mm$.}
\end{figure}

The transmission spectrum of the Nd$^{3+}$:YVO$_{4}$ FADOF has two contributions: one is the
absorption of the Nd$^{3+}$ ions and the other is the difference in their refractive indices. Hence, knowledge about the
absorption properties of the Nd$^{3+}$ ions and the complex refractive indices enable us to calculate the entire transmission
of the Nd$^{3+}$:YVO$_{4}$ FADOF.

In this work, we utilize the strong $^{4}I_{9/2}\leftrightarrow{ }^4F_{3/2}$ transition of Nd$^{3+}$ around 879.7 nm in a Nd$^{3+}$:YVO$_{4}$
crystal (doping level, 10 ppm.); which has been systematically investigated as a
candidate for quantum memory  \cite{S.R.Hasting-Simon2008,Hugues de Riedmatten2008,Mikael Afzelius2010}.
The YVO$_{4}$ crystal is uniaxial and Nd$^{3+}$ ions substitute for Y$^{3+}$ ions in site of D$_{2d}$
point symmetry. In a crystal field of this symmetry, the ground state $^{4}I_{9/2}$ splits into five Kramers doublets  ($Z_{1}$  -  $Z_{5}$)  and the
${ }^4F_{3/2}$ excited state splits into two Kramers doublets  ($R_{1}$  and  $R_{2}$).   Since the transition between $Z_{1}$ and $R_{1}$
possesses the largest oscillator strength, the other transitions can be reasonably ignored \cite{O.Guiilot-Noel2000,V Mehta2000,Y.Sun2002}. Under a homogeneous, constant magnetic field, the
ground and excited state doublets split into two spin sublevels due to a strong first-order Zeeman interaction. This gives rise to a four-level system
with two ground levels and two excited levels, and the four associated possible transitions hereafter are denoted as $a, b, c$ and $d$.
The energy levels are shown in Fig. 2(a). Here, $P_{\shortparallel}$ and $P_{\perp}$, respectively, denote the transition dipole operators parallel and perpendicular
to the $c$ axis of the Nd$^{3+}$:YVO$_{4}$ crystal, and $|\Psi_{\pm\mu}\rangle$ denotes the wave functions of the states. The selection rules which hold for electric
dipole transitions between $|\Psi_{\pm\mu}\rangle$ are given by \cite{Mikael Afzelius2010}.
\begin{eqnarray}
\langle\Psi_{\mu}|P_\shortparallel|\Psi_{\mu'}\rangle&=&0\ , \ \ \ \ \ \ \    mod\ (\ \mu-\mu',2\ )=0.\nonumber \\
\langle\Psi_{\mu}|P_{\perp}|\Psi_{\mu'}\rangle&=&0\ , \ \ \ \ \ \ \      mod\ (\ \mu-\mu',2\ )=1.
\end{eqnarray}
Hence, the transitions $b, c$ and $a, d$ are linearly polarized light along the $H$ and $V$ direction, respectively.

We ignore the coherence between the energy levels considered and treat each ion as consisting of four two-level systems in the Nd$^{3+}$:YVO$_{4}$ crystal.
Thus, we can approximate the complex susceptibility along the H- and V-direction $\chi_{\scriptscriptstyle H/V}(\omega)$ as \cite{Michael Fleischhauer2005,Chuan-Feng Li2011,Ryan M. Camacho2007}
\begin{eqnarray}
\chi_{\scriptscriptstyle H}(\omega)=-\sum_{q=a}^{d} \frac{\alpha_{\scriptscriptstyle qH}\delta_{\scriptscriptstyle qH}}{k_{\scriptscriptstyle H0}(\omega)[(\omega-\omega_{\scriptscriptstyle q})+i\delta_{\scriptscriptstyle qH}]}\ \ ,\\
\chi_{\scriptscriptstyle V}(\omega)=-\sum_{q=a}^{d} \frac{\alpha_{\scriptscriptstyle qV}\delta_{\scriptscriptstyle qV}}{k_{\scriptscriptstyle V0}(\omega)[(\omega-\omega_{\scriptscriptstyle q})+i\delta_{\scriptscriptstyle qV}]}\ \ ,
\end{eqnarray}
where $q$ can substitute for $a, b, c$ and $d$ according to the transition lines; $\alpha_{\scriptscriptstyle qH/V}$ is the on-resonance absorption coefficient;
$k_{\scriptscriptstyle H0/V0}(\omega)=n_{\scriptscriptstyle H0/V0}(\omega)\ \omega/c$, where $n_{\scriptscriptstyle H0/V0}$ is the intrinsic refractive
index in the Nd$^{3+}$:YVO$_{4}$ crystal, which in turn can be derived from the Snellmeier Equation \cite{Sellmeier equations}; $c$ is the speed of light in vacuum;
$2\delta_{\scriptscriptstyle qH/V}$ is the absorption bandwidth of the transition line; $\omega_{\scriptscriptstyle q}$ is the resonance frequency; and
$\omega$ is the incident light frequency.
The complex susceptibility $\chi_{\scriptscriptstyle H/V}(\omega)$ relates the complex wave number $k_{\scriptscriptstyle H/V}(\omega)$
and frequency $\omega$ in the dispersion relation $k_{\scriptscriptstyle H/V}(\omega)=k_{\scriptscriptstyle H0/V0}(\omega)\sqrt{1+\chi_{\scriptscriptstyle H/V}(\omega)}$. For
$\chi_{\scriptscriptstyle H/V}(\omega)\ll1$, the relationship can be approximated as
\begin{eqnarray}
k_{\scriptscriptstyle H/V}(\omega)\approx k_{\scriptscriptstyle H0/V0}(\omega)\big[1+\frac{1}{2}\chi_{\scriptscriptstyle H/V}(\omega)\big] \ .
\end{eqnarray}
\begin{figure}[tbph]
\begin{center}
\includegraphics[width= 3in]{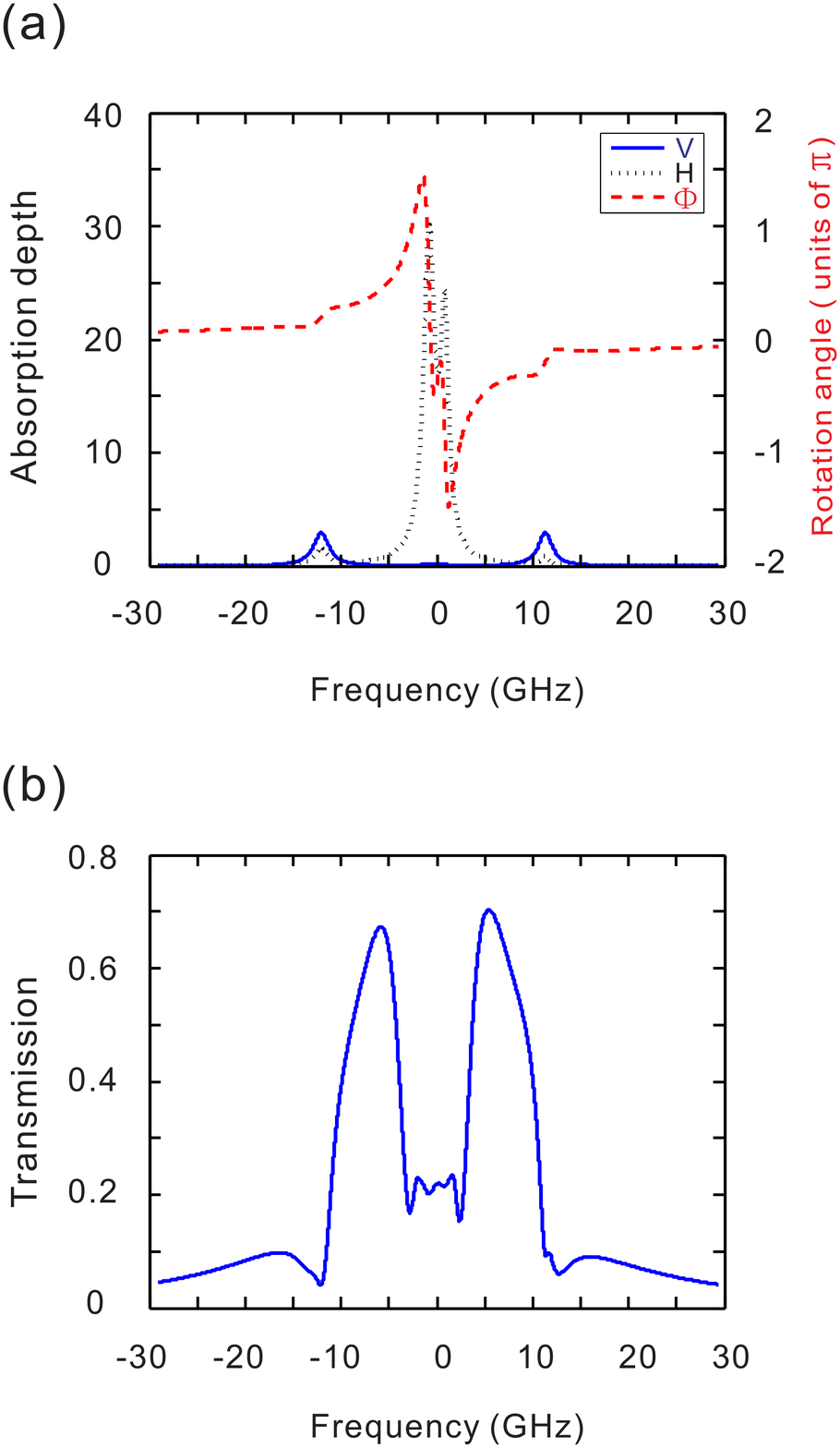}
\end{center}
\caption{(Color online)    (a) The simulation results of the absorption depth $d_{\scriptscriptstyle H/V}(\Delta)$ (the black dotted line for $H$-polarized light and the blue solid line for $V$-polarized light) and the Faraday rotation
angle $\Phi(\Delta)$ (the red dashed line) in the Nd$^{3+}$:YVO$_{4}$ FADOF.
(b) The simulation results of the transmission $T(\Delta)$ in the Nd$^{3+}$:YVO$_{4}$ FADOF, which predict a maximum transmission efficiency of $71\%$ and a double-peaked
transmission spectrum with a bandwidth of 6 GHz. }
\end{figure}

The absorption depth $d(\omega)$ is defined in terms of intensity attenuation $exp\ [-d(\omega)]$ through the sample. Hence,
\begin{eqnarray}
d(\omega)=\alpha(\omega) L=2Im\big[k_{\scriptscriptstyle H/V}(\omega)\big]L \ \ .
\end{eqnarray}

We substitute Equations (7), (8), and (9) into Eq. (10) to simulate the absorption depth for both $H$- and $V$- polarized incident light. All the parameters are based on the experiment results \cite{S.R.Hasting-Simon2008,Mikael Afzelius2010}.
The simulation results are shown in Figs .2(b) and (c), which are in good agreement with the experimental results.

From the relationship $n_{\scriptscriptstyle H/V}(\omega)=k_{\scriptscriptstyle H/V}(\omega)c/\omega$, where $n_{\scriptscriptstyle H/V}(\omega)$
is the refractive index along the $H$ and $V$ direction, we obtained the dispersion curve in this system: this contains both normal and anomalous dispersion parts.
Hence, the value of $n_{\scriptscriptstyle H/V}(\omega)$ is strongly dependent on the frequency of the incident light, which
leads to a polarization rotation for light passing through the crystals. If the incident light is outside
the absorption bands and the rotation angle approximates to odd multiples of $\pi/2$, it will pass through the two crossed polarizers. We calculate the frequency-dependent
Nd$^{3+}$:YVO$_{4}$ FADOF rotation angle $\Phi(\Delta)$ [Eq. (5)], absorption depth $d_{\scriptscriptstyle H/V}(\Delta)$ [Eq.( 10)] and transmission $T(\Delta)$ [Eq. (4)], where $\Delta=\omega-\omega_{b}$,
by changing the magnetic field strength and the crystal length. Note that the splitting of energy levels is approximately linear with the magnetic field \cite{S.R.Hasting-Simon2008}.
We predict a maximum transmission efficiency of $71\%$ and a double-peaked transmission spectrum with a bandwidth of 6 GHz under the condition that the magnetic $B=0.775\ T$ and the
crystal length $L=8.5\ mm$. The simulation results are shown in Fig.3. We also calculated the transmission $T(\Delta)$ under different magnetic fields: the
conclusion is that both the transmission and bandwidth increase with the strength of the magnetic field. In other words, there is a trade-off between the transmission and the bandwidth.

In summary, we have proposed the scheme of a solid-state Faraday anomalous dispersion optical filter. A theoretical model of the solid-
state FADOF's transmission has been presented. We simulate the performance of the Nd$^{3+}$:YVO$_{4}$ FADOF, which works with
the strong $^{4}I_{9/2}\leftrightarrow{ }^4F_{3/2}$ transition of Nd$^{3+}$ around 879.7 nm, based on previous experimental measurements
of the absorption properties of the Nd$^{3+}$:YVO$_{4}$ crystal. The maximum transmission efficiency can attain $71\%$ with a double-peaked
transmission spectrum, whose bandwidth is 6 GHz.

As an ultra-narrow bandwidth filter, a FADOF performed in a solid-state device is very appealing. Apart
from the Nd$^{3+}$ doped YVO$_{4}$ crystal, there are varieties of rare-earth doped solid materials which have been recently demonstrated to play key
roles in quantum information: the working elements include Er$^{3+}$ \cite{Bjorn Lauritzen2010}, Eu$^{3+}$ \cite{Alexander AL2006},
Pr$^{3+}$ \cite{Hedges Morgan P.2010}, Tm$^{3+}$ \cite{Saglamyurek Erhan2011} and others. Similar to the Nd$^{3+}$:YVO$_{4}$ crystal, these materials
may possess potential properties to construct more solid-state FADOFs for different applications. Specifically,  the Er$^{3+}$ doped crystals
have a transition around 1530 nm between the ground state $^{4}I_{15/2}$ and the excited state $^{4}I_{13/2}$ \cite{Thomas Bottger2006}.
It offers the possibility to construct ultra-narrow-bandwidth filters that work at telecommunication wavelengths, which is difficult for the conventional FADOFs based on atomic vapors to
achieve. Furthermore, solids doped with rare-earth ions are unique physical systems in which large ensembles of atoms are naturally trapped in a solid-state matrix, which inhibits the
Doppler effect due to the motion of the atoms. These systems are stable, commercially available, and easily integrated. For these reasons, our proposal may have important applications in
optical communications.

This research is supported by the National Basic Research Program (2011CB921200) and National
Natural Science Foundation of China (Grant No. 60921091 and 10874162).

\end{document}